\documentclass[11pt]{article}

\usepackage[utf8]{inputenc}
\usepackage[margin=1in]{geometry}
\usepackage{amsmath,amssymb}
\usepackage{graphicx}
\usepackage{booktabs}
\usepackage{caption}
\usepackage{array}
\usepackage[affil-it]{authblk}
\usepackage{url}
\usepackage[colorlinks=true,linkcolor=blue,citecolor=blue,urlcolor=blue]{hyperref}

\newcolumntype{L}[1]{>{\raggedright\arraybackslash}p{#1}}
\newcolumntype{C}[1]{>{\centering\arraybackslash}p{#1}}

\title{\textbf{Attributing Sensor Deviations to Degradation, Weather, or Attack in Oilfield Digital Twins: A Simulation Study of Probabilistic Attribution and Cost-Based Decisions}}

\author[1]{Mustafa S. Aljumaily}
\author[2]{Nawar S. Alseelawi}
\author[3]{Hayder Kareem Abed}
\affil[1]{Research and Development (R\&D) Department, Daw Alfada Company, Baghdad, Iraq\\ \texttt{mustafa.s@daw-alfada.com}}
\affil[2]{University of Misan, Maysan, Iraq\\ \texttt{nawar.alseelawi@uomisan.edu.iq}}
\affil[3]{CEO, Daw Alfada Company, Baghdad, Iraq\\ \texttt{haider.k@daw-alfada.com}}
\date{}

\begin{document}

\maketitle

\begin{abstract}
\noindent
When an oilfield digital twin disagrees with its instruments, the operator must decide whether the cause is hardware degradation, harsh-weather effects, or malicious data manipulation. Existing digital-twin work appears to treat these causes separately: sensor-validation architectures target faults, and attack-focused twins have been evaluated on water-sector testbeds. We study joint cause attribution on a simulated four-well wellpad with 15 coupled instruments, legitimate operating transients, and weather. A physics-informed twin, identified from normal data only, produces analytical-redundancy residuals; windowed features feed a gradient-boosted classifier whose posterior drives an alarm gate with a fixed false-alarm rate and a cost-based decision rule that may defer to an analyst. On three independently generated sites, the classifier reaches macro-F1 of 0.818 on windows where the injected deviation is observable (0.723 when latent post-onset windows are included at the primary site). The physics twin accounts for essentially all of this: removing the data-driven twin changes macro-F1 by less than 0.01, whereas removing all twins drops it to about 0.58. Attacks are detected quickly (median 1.7~h) but attributed correctly at alarm time only 42\% of the time, rising to 73\% six hours later. A cost-aware policy that defers ambiguous cases had the lowest expected cost among all policies in every one of 144 cost and prior settings tested, sometimes by a small margin; the result depends on illustrative costs and on analysts resolving deferred cases. A twin-aware attacker was detected in 45\% of episodes yet almost never attributed to attack. These results are conditional on the simulator's generative assumptions and have not been validated on field data.

\end{abstract}

\section{Introduction}\label{sec:intro}

A digital twin of an oil-and-gas asset is only as trustworthy as the instruments feeding it. In the field, those instruments face heat, dust, vibration, and long maintenance intervals, and, increasingly, network-connected gateways that an adversary may compromise. When the twin and an instrument disagree, three explanations compete: the instrument is degrading, the environment is distorting it, or someone is manipulating the value. The right response differs sharply. Degradation calls for a maintenance visit, weather effects for compensation or shielding, and manipulation for incident response and isolation of the affected gateway. Choosing wrongly is costly in both directions: dismissing an attack as wear leaves it running, and escalating every drifting transmitter to the security team exhausts analysts.

Prior work on digital-twin sensor validation and on attack detection has largely proceeded on separate tracks. To our reading, sensor fault detection, isolation and accommodation (SFDIA) for industrial twins \cite{darvishi2023,darvishi2021} addresses sensor failure, while attack-discriminating twins \cite{homaei2026} address malicious manipulation and have been evaluated on water-treatment and water-distribution testbeds \cite{goh2016,ahmed2017wadi}. We are not aware of work that jointly attributes a twin--instrument deviation to degradation, weather, or attack under oilfield-like conditions, although our literature search was brief and this claim should be checked before it is relied upon.

This paper makes four contributions, each limited to a simulation setting. First, we define the three-way attribution problem and an open simulator for it, including legitimate operating transients that a naive detector would flag. Second, we evaluate a physics-informed twin, identified from normal data only, as the source of residual features, against a data-driven twin and against no twin. Third, we report posterior reliability, detection delay and attribution quality as a function of time since alarm, and we show that a cost-aware decision rule which can defer to an analyst outperforms both argmax attribution and the `escalate everything' baseline under a wide range of cost settings. Fourth, we report where the approach fails, including single-sensor faults versus attacks that are indistinguishable by construction and a twin-aware attacker.

We emphasize what this paper does not show. No field data were used. The simulator's assumptions determine what is identifiable, the cost matrix is illustrative, and every conclusion is conditional on both.

\section{Background}\label{sec:background}

Model-based fault diagnosis builds residuals from analytical redundancy among measured variables and isolates faults by their signature across residuals \cite{isermann2005}. Digital-twin variants combine physics with learned models; a recurrent graph-convolutional architecture for sensor fault detection, isolation and accommodation was reported for industrial twins \cite{darvishi2023}, following an earlier real-time scheme \cite{darvishi2021}. On the security side, false data injection can evade residual-based detectors when the injected vector lies in the range of the measurement model \cite{liu2011fdi}, and replay attacks can be countered by exploiting the noise fingerprint of genuine measurements \cite{mo2009replay}. A recent self-defending twin discriminates single- and multi-stage attacks on SWaT and WADI \cite{homaei2026,goh2016,ahmed2017wadi}. Our `twin-aware' stress test follows the undetectability logic of \cite{liu2011fdi} but is deliberately imperfect (Section~\ref{sec:limitations}).

For decision-making, we use gradient-boosted trees \cite{pedregosa2011} with isotonic recalibration \cite{niculescu2005,guo2017} and evaluate calibration by expected calibration error (ECE) with ten bins.

\section{Problem formulation and threat model}\label{sec:problem}

At each time step an operator observes 15 instrument readings $y_t$ and a weather-station record (ambient temperature and a dust index). A persistent deviation may begin at an unknown onset. The task is to (i) raise an alarm at a controlled false-alarm rate, (ii) produce a calibrated posterior over \{normal, fault, environment, attack\}, and (iii) choose an action from \{ignore, maintenance, compensate, escalate, analyst review\} minimizing expected cost.

The attacker can alter sensor values by compromising a gateway (several sensors on the same RTU) or a single field device. The weather station is assumed trustworthy. Actuator manipulation, network-layer attacks, denial of service, and attacks on the twin itself are out of scope.

\section{Simulated wellpad}\label{sec:sim}

The simulator (open code accompanies this paper) represents four wells feeding a header, with 15 instruments (wellhead pressure, temperature and flow per well; header pressure, temperature and flow) served by three gateways. The true process satisfies mass balance, a quadratic choke/flowline pressure--flow relation, a header pressure--flow relation, and a first-order thermal lag at each wellhead. Legitimate operating changes (choke steps, short shut-ins, slugging) and weather (diurnal ambient temperature, random dust events) are always present, so `something changed' is not evidence of an anomaly. Episodes last 72~h at 10-minute sampling; an anomaly, when present, starts between 25\% and 60\% of the episode and persists.

\begin{table}[ht]
\centering
\caption{Generative model of the four causes.}
\label{tab:causes}
\small
\begin{tabular}{L{2.0cm}L{3.6cm}L{3.6cm}L{4.4cm}}
\toprule
\textbf{Cause} & \textbf{Modes} & \textbf{Affected sensors} & \textbf{Distinguishing structure} \\
\midrule
Normal & none & none & operating transients and weather present \\
Fault & drift, stuck, step, gain, noise bursts & 1 sensor (85\%) or 2, independent & sensor-local; independent of weather \\
Environment & heat, dust, both & 2--8 sensors & coupled to ambient temperature / dust index \\
Attack & bias, ramp, replay, scale, mass-balance-preserving & 70\%: 2+ sensors on one gateway; 30\%: one device & abrupt or coordinated; replay repeats the noise pattern \\
Stress test only & twin-aware (`full') & 7 sensors & keeps all physics residuals near zero \\
\bottomrule
\end{tabular}
\end{table}

Fault and attack magnitudes are drawn from the same log-uniform range (0.5--8 units, a unit being about three times instrument accuracy), and single-sensor drift or step faults are statistically similar to single-sensor ramp or bias attacks. These cases are therefore partly indistinguishable by construction; the classifier's confusion there reflects the simulator, not only the method.

\section{Method}\label{sec:method}

\begin{figure}[ht]
\centering
\includegraphics[width=0.95\linewidth]{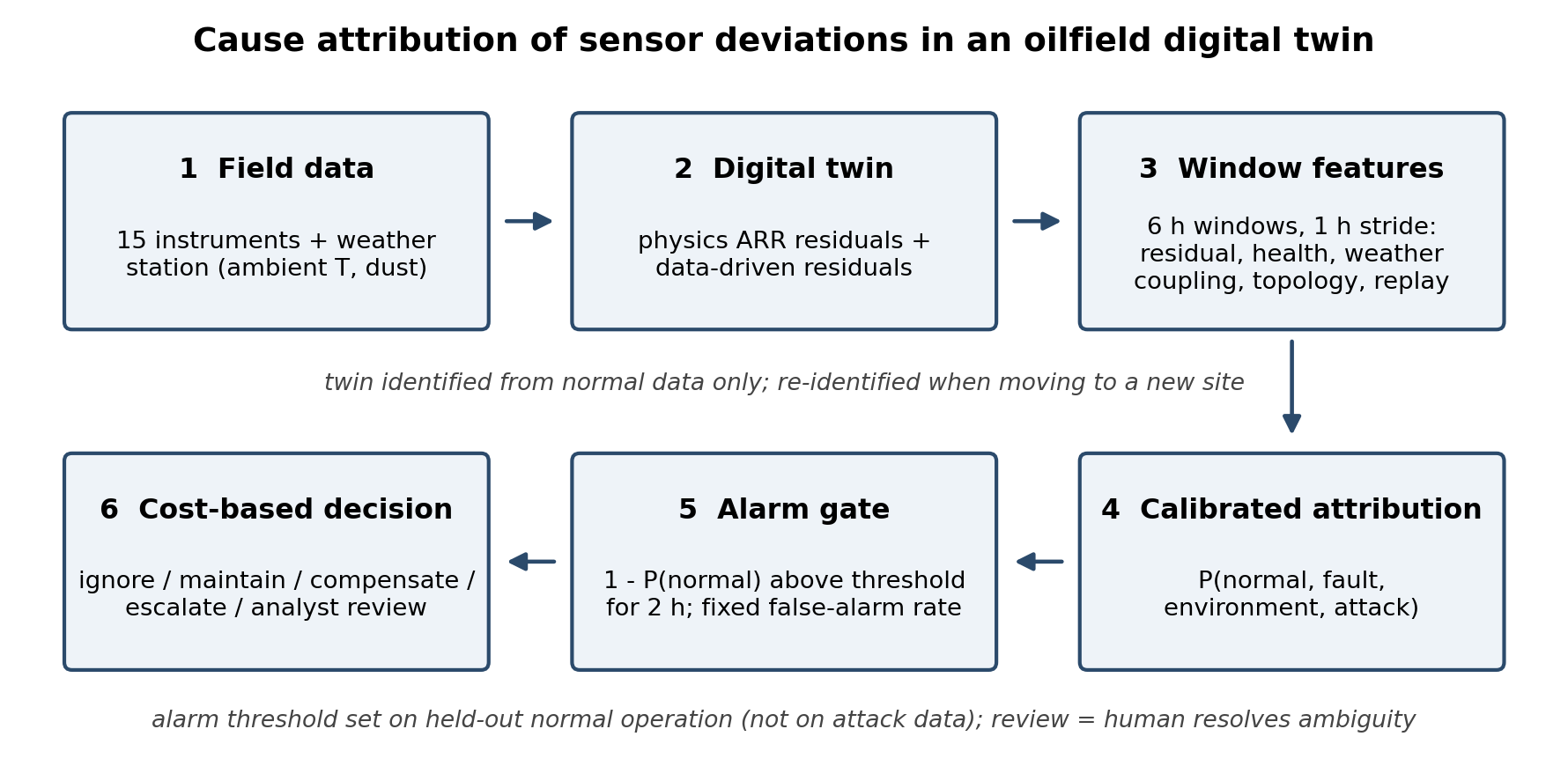}
\caption{Processing pipeline. The twin is identified from normal data only; the alarm threshold is set on held-out normal operation.}
\label{fig:pipeline}
\end{figure}

\subsection{Twins and residuals}\label{sec:twins}

The physics twin evaluates twelve analytical-redundancy relations (ARRs): header mass balance; one choke relation per well (pressure drop against squared flow); one wellhead temperature relation per well as a function of flow and ambient temperature; header temperature against the flow-weighted well temperature; and header pressure against total flow computed two ways. Functional forms come from process physics; parameters are identified from normal data only, with no simulator internals. Residuals are standardized with robust statistics from normal data. The data-driven twin predicts each sensor from all others and ambient temperature using gradient boosting trained on normal data, and its residuals are standardized the same way.

\subsection{Features}\label{sec:features}

Six-hour windows with a one-hour stride yield 494 features in eight groups (the proposed model uses seven groups, 382 features, and leaves out the raw-signal statistics that serve the no-twin baseline): residual statistics (arr, dd), z-scored raw-signal statistics (raw), instrument health (noise ratio and flatness), coupling of residuals to weather covariates (cov), sensor-isolation and gateway-level aggregates (topo), a replay fingerprint (maximum lagged correlation of a sensor's high-passed signal with earlier history), and operating context. Permutation-invariant aggregates (maxima, counts, per-gateway maxima) accompany per-sensor features.

\subsection{Attribution, alarm gate and decision}\label{sec:decision}

A gradient-boosted classifier outputs class probabilities, recalibrated by isotonic regression on a separate calibration set. An alarm is raised when $1 - P(\text{normal})$ exceeds a threshold for two consecutive windows; the threshold is set on 150 held-out normal episodes for 0.05 false alarms per day. Given the posterior $p$ at alarm time, the policy chooses the action minimizing $\sum_c p_c \cdot \mathrm{Cost}(c, \text{action})$, with analyst review as an action of fixed cost. We also evaluate deciding six hours after the alarm using the mean posterior over that interval.

\begin{table}[ht]
\centering
\caption{Illustrative operator cost matrix (rows: true cause; columns: action). Review resolves an attack with probability 0.95 (cost $3 + 0.05 \times 60$).}
\label{tab:cost}
\begin{tabular}{lccccc}
\toprule
\textbf{True cause} & \textbf{Ignore} & \textbf{Maintain} & \textbf{Compensate} & \textbf{Escalate} & \textbf{Review} \\
\midrule
Normal      & 0  & 4  & 2  & 6 & 3 \\
Fault       & 10 & 1  & 3  & 8 & 3 \\
Environment & 6  & 4  & 1  & 6 & 3 \\
Attack      & 60 & 50 & 50 & 2 & 6 \\
\bottomrule
\end{tabular}
\end{table}

\section{Experimental protocol}\label{sec:protocol}

For each of three independently generated sites (different wells, climate and noise), we generate disjoint sets: 250 normal episodes to identify twins, 150 normal episodes to set the alarm threshold, 800 mixed episodes to train classifiers, 250 to calibrate probabilities, and 500 for testing. Class proportions in generated episodes are 30\% normal, 25\% fault, 22\% environment and 23\% attack. Baselines and ablations use the same splits and the same false-alarm target.

\paragraph{Window labels.} A post-onset window is `manifest' if some sensor's RMS injected deviation within the window exceeds 1.5 instrument-noise standard deviations, a criterion fixed before results were inspected. Non-manifest post-onset windows are excluded from training and from the primary window metric, because no method could recover their label; we report results on all post-onset windows alongside. Detection delay is always measured from the true onset.

\paragraph{Methods compared.} Proposed (all feature groups); physics twin only; data-driven twin only; no twin (raw signals only); four leave-one-group-out ablations; and four detector baselines that share the alarm gate but cannot attribute: a generic anomaly detector followed by always-maintain or always-escalate, an attack-only detector trained on normal versus attack data, and a fault-only detector. Confidence intervals are episode-level bootstrap intervals (500 resamples for window metrics, 200 for streaming metrics).

\paragraph{Deviations from a purely pre-specified protocol.} Permutation-invariant aggregate features were added after an initial prototype run showed that replay attacks, which can hit any of 15 sensors, were learned poorly from per-sensor features (replay recall 0.47 rose to 0.75 on the prototype). This design decision was made on the same simulator, although evaluation used freshly generated data. The cost matrix, the manifest threshold and the false-alarm target were fixed before results.

\section{Results}\label{sec:results}

\subsection{Attribution quality and calibration}\label{sec:results-attr}

\begin{figure}[ht]
\centering
\includegraphics[width=0.95\linewidth]{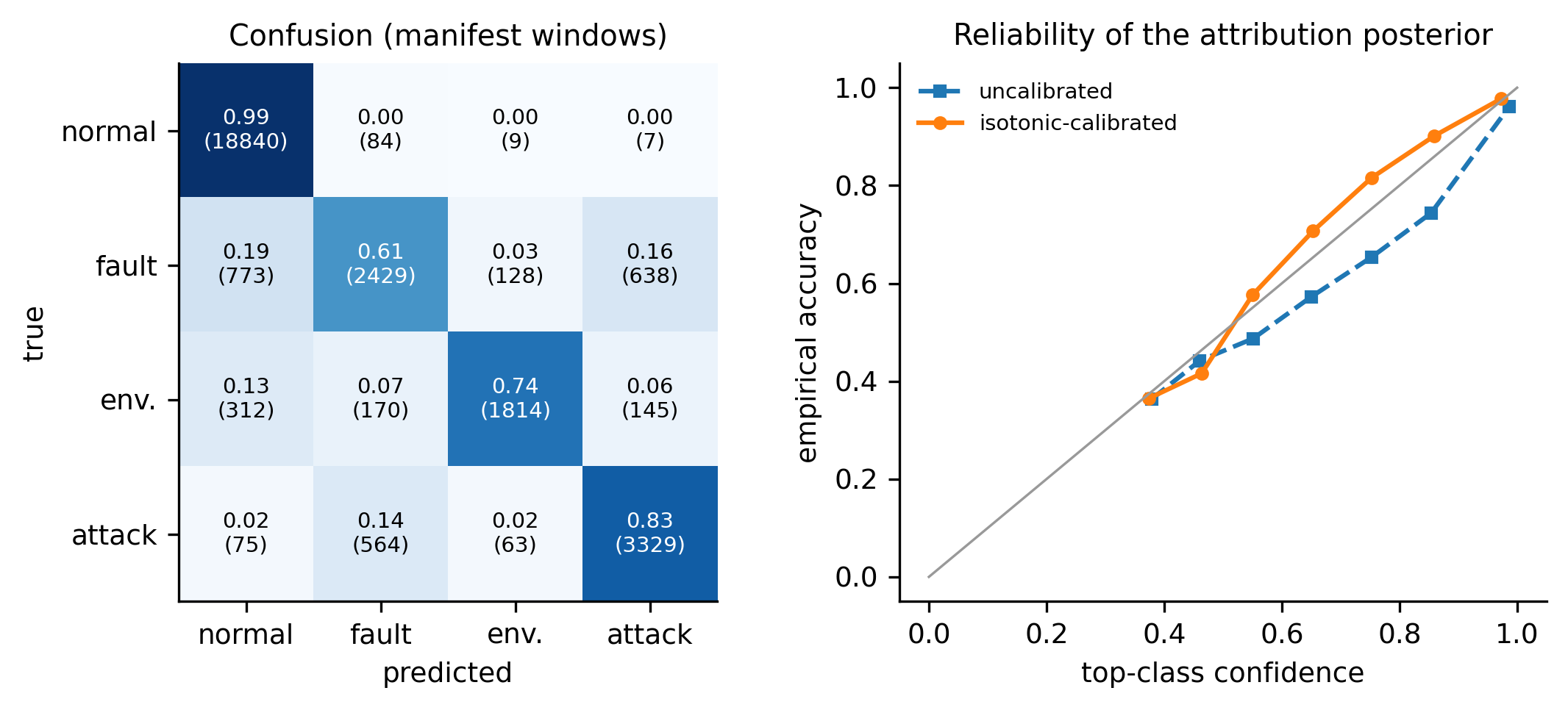}
\caption{Left: row-normalized confusion matrix on manifest test windows, primary site (counts in parentheses). Right: reliability of the top-class posterior at the primary site, before and after isotonic recalibration.}
\label{fig:confusion}
\end{figure}

On manifest windows the proposed model reaches accuracy 0.899 and macro-F1 0.818 [0.794, 0.841] at the primary site, and 0.818, 0.819, 0.816 at the three sites. Per-class F1 at the primary site is 0.97 / 0.67 / 0.81 / 0.82 for normal, fault, environment and attack. Including latent post-onset windows lowers macro-F1 to 0.723 [0.695, 0.748]. The dominant error is the fault--attack confusion (Figure~\ref{fig:confusion}, primary site): about 16\% of fault windows are labelled attack and 14\% of attack windows fault, and 19\% of manifest fault windows are labelled normal.

Recalibration did not help consistently. ECE fell from 0.036 to 0.021 at the primary site (Figure~\ref{fig:confusion}, right), but rose from 0.032 to 0.064 and from 0.029 to 0.042 at the other two sites; at the primary site the recalibrated posterior is under-confident in the 0.6--0.9 band. The posterior is moderately reliable in every case (ECE 0.021--0.064), but our data do not support the claim that isotonic recalibration is necessary.

\subsection{What contributes: ablations}\label{sec:results-abl}

\begin{table}[ht]
\centering
\caption{Window-level macro-F1 on held-out windows. Primary site with 95\% episode-bootstrap CI, mean over the three sites, and the paired difference from the full model pooled over the three sites (paired episode bootstrap, 2000 resamples).}
\label{tab:ablation}
\small\setlength{\tabcolsep}{4pt}
\begin{tabular}{L{3.0cm}C{2.9cm}C{1.9cm}C{1.5cm}C{3.4cm}}
\toprule
\textbf{Variant} & \textbf{Manifest windows, primary site [95\% CI]} & \textbf{All post-onset windows} & \textbf{Mean of 3 sites} & \textbf{$\Delta$ vs proposed, pooled [paired 95\% CI]} \\
\midrule
Proposed (all groups)  & 0.818 [0.794, 0.841] & 0.723 & 0.818 & --- \\
Physics twin only      & 0.816 [0.791, 0.837] & 0.722 & 0.820 & $+0.002$ [$-0.002$, $+0.007$] \\
Data-driven twin only  & 0.772 [0.747, 0.796] & 0.678 & 0.768 & $-0.050$ [$-0.058$, $-0.041$] \\
No twin (raw signals)  & 0.592 [0.558, 0.624] & 0.521 & 0.579 & $-0.239$ [$-0.256$, $-0.223$] \\
$-$ weather coupling     & 0.797 [0.768, 0.819] & 0.707 & 0.797 & $-0.020$ [$-0.026$, $-0.015$] \\
$-$ topology / isolation & 0.814 [0.788, 0.838] & 0.719 & 0.811 & $-0.007$ [$-0.011$, $-0.003$] \\
$-$ replay fingerprint   & 0.804 [0.780, 0.825] & 0.709 & 0.805 & $-0.012$ [$-0.018$, $-0.008$] \\
$-$ instrument health    & 0.798 [0.774, 0.820] & 0.704 & 0.797 & $-0.020$ [$-0.028$, $-0.013$] \\
\bottomrule
\end{tabular}
\end{table}

The physics twin carries almost all of the signal: the physics-only model differs from the full model by $+0.002$ [$-0.002$, $+0.007$] in macro-F1 (variant minus full, pooled over sites); the paired interval includes zero, so the data-driven twin's marginal contribution is indistinguishable from nothing here, to within about $\pm 0.007$. The data-driven twin alone is 0.050 lower than the full model, and without any twin macro-F1 falls by 0.239 to 0.579. Each remaining group contributes a small but reliable amount: removing weather coupling costs 0.020, instrument health 0.020, the replay fingerprint 0.012 and topology features 0.007, with paired intervals excluding zero and the same sign at all three sites. These intervals reflect sampling variability in the test episodes; they do not capture uncertainty about the simulator's assumptions, and the three sites share one generative process. Weather coupling matters most for the environment class (F1 0.81 versus 0.75 without it at the primary site).

\begin{figure}[ht]
\centering
\includegraphics[width=0.8\linewidth]{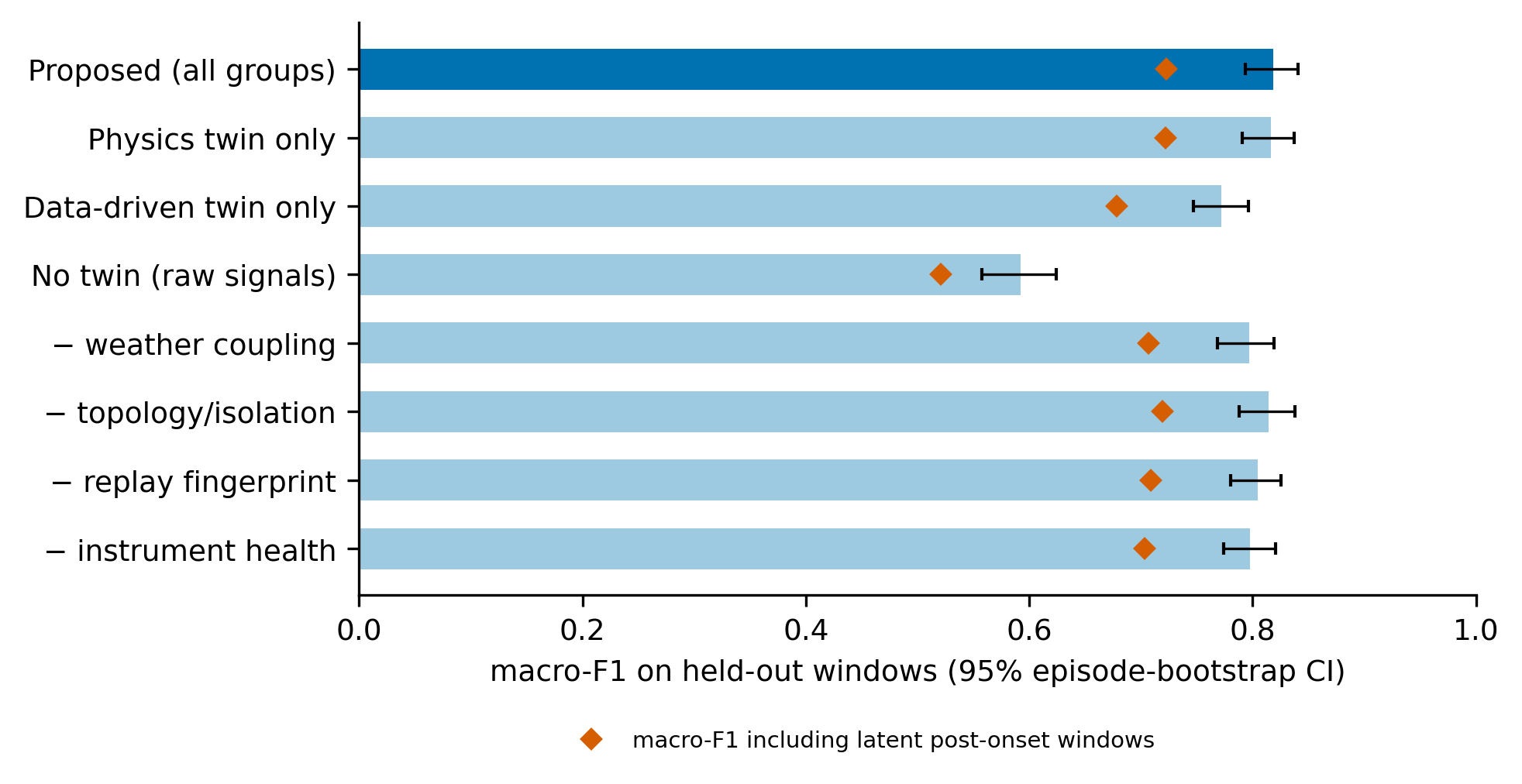}
\caption{Macro-F1 by feature configuration at the primary site. Bars: manifest windows with 95\% episode-bootstrap CI; diamonds: including latent post-onset windows.}
\label{fig:ablation}
\end{figure}

Per-mode recall on manifest windows (mean over sites) is highest for stuck sensors (1.00), mass-balance-preserving attacks (1.00) and replay (0.95), and lowest for drift (0.48), noise bursts (0.54), steps (0.48), gain errors (0.53) and ramps (0.57); see Appendix~\ref{app:recall}.

\subsection{Streaming detection, attribution over time, and operator cost}\label{sec:results-stream}

At the alarm threshold set for 0.05 false alarms per day, the realized rate on test normal episodes was 0.027 per day at the primary site. Median detection delay from true onset was 7.3~h for faults, 8.2~h for environment effects and 1.7~h for attacks; the 90th percentiles were 28.7, 23.2 and 8.0~h. Detection rates were 91\%, 96\% and 100\%. Attribution improves with waiting (Table~\ref{tab:bycause}).

\begin{table}[ht]
\centering
\caption{Detection and attribution by cause (argmax attribution), primary site; attribution ranges over the three sites in parentheses.}
\label{tab:bycause}
\small
\begin{tabular}{lcccc}
\toprule
\textbf{Cause} & \textbf{Detected} & \textbf{Median delay (h)} & \textbf{Correct at alarm} & \textbf{Correct 6 h later} \\
\midrule
Fault       & 91\%  & 7.3 & 0.74 (0.59--0.74) & 0.79 (0.68--0.79) \\
Environment & 96\%  & 8.2 & 0.81 (0.80--0.89) & 0.86 (0.83--0.88) \\
Attack      & 100\% & 1.7 & 0.42 (0.35--0.45) & 0.73 (0.68--0.77) \\
\bottomrule
\end{tabular}
\end{table}

Attacks are detected fastest but attributed worst at the moment of alarm (42\% correct), because the alarm fires after only two windows and early evidence resembles a fault. Six hours later, 73\% are attributed correctly. Fault and environment attribution changes little with waiting.

\begin{table}[ht]
\centering
\caption{Operator cost and attack handling, primary site (500 test episodes). Cost is the mean over episodes under Table~\ref{tab:cost}, with equal class weights and with an illustrative deployment prior (88\% normal, 5\% fault, 5\% environment, 2\% attack). All methods share the alarm-gate false-alarm target. For policies with an analyst-review action, attacks not escalated or treated as benign were sent to review.}
\label{tab:opcost}
\footnotesize\setlength{\tabcolsep}{3.5pt}
\begin{tabular}{L{4.3cm}C{2.3cm}C{1.8cm}C{1.5cm}C{2.2cm}C{1.7cm}}
\toprule
\textbf{Method} & \textbf{Cost, equal weights [95\% CI]} & \textbf{Cost, deployment prior} & \textbf{Attacks escalated} & \textbf{Attacks missed or treated as benign} & \textbf{Non-attacks escalated} \\
\midrule
Proposed (argmax)                              & 8.80 [7.59, 9.95]   & 1.34 & 0.42 & 0.58 & 0.05 \\
Proposed (min-cost)                            & 3.74 [3.10, 4.46]   & 0.95 & 0.89 & 0.11 & 0.18 \\
Proposed (min-cost + review)                   & 3.18 [2.82, 3.54]   & 0.81 & 0.29 & 0.04 & 0.02 \\
Proposed (min-cost + review, decide 6 h later) & 2.56 [2.34, 2.83]   & 0.74 & 0.72 & 0.01 & 0.03 \\
Physics twin only (argmax)                     & 8.37 [7.26, 9.46]   & 1.27 & 0.46 & 0.54 & 0.06 \\
Data-driven twin only (argmax)                 & 8.09 [7.02, 9.32]   & 1.21 & 0.50 & 0.50 & 0.08 \\
No twin, raw signals (argmax)                  & 8.82 [7.73, 10.30]  & 1.42 & 0.60 & 0.40 & 0.08 \\
Generic detector $\to$ maintenance             & 14.12 [14.01, 14.25] & 1.85 & 0.00 & 1.00 & 0.00 \\
Generic detector $\to$ escalate                & 4.28 [4.21, 4.38]   & 1.59 & 1.00 & 0.00 & 0.64 \\
Attack-only detector $\to$ escalate            & 4.45 [4.25, 4.71]   & 1.54 & 0.99 & 0.01 & 0.59 \\
Fault-only detector $\to$ maintenance          & 14.20 [14.06, 14.33] & 1.76 & 0.00 & 1.00 & 0.00 \\
\bottomrule
\end{tabular}
\end{table}

\begin{figure}[ht]
\centering
\includegraphics[width=\linewidth]{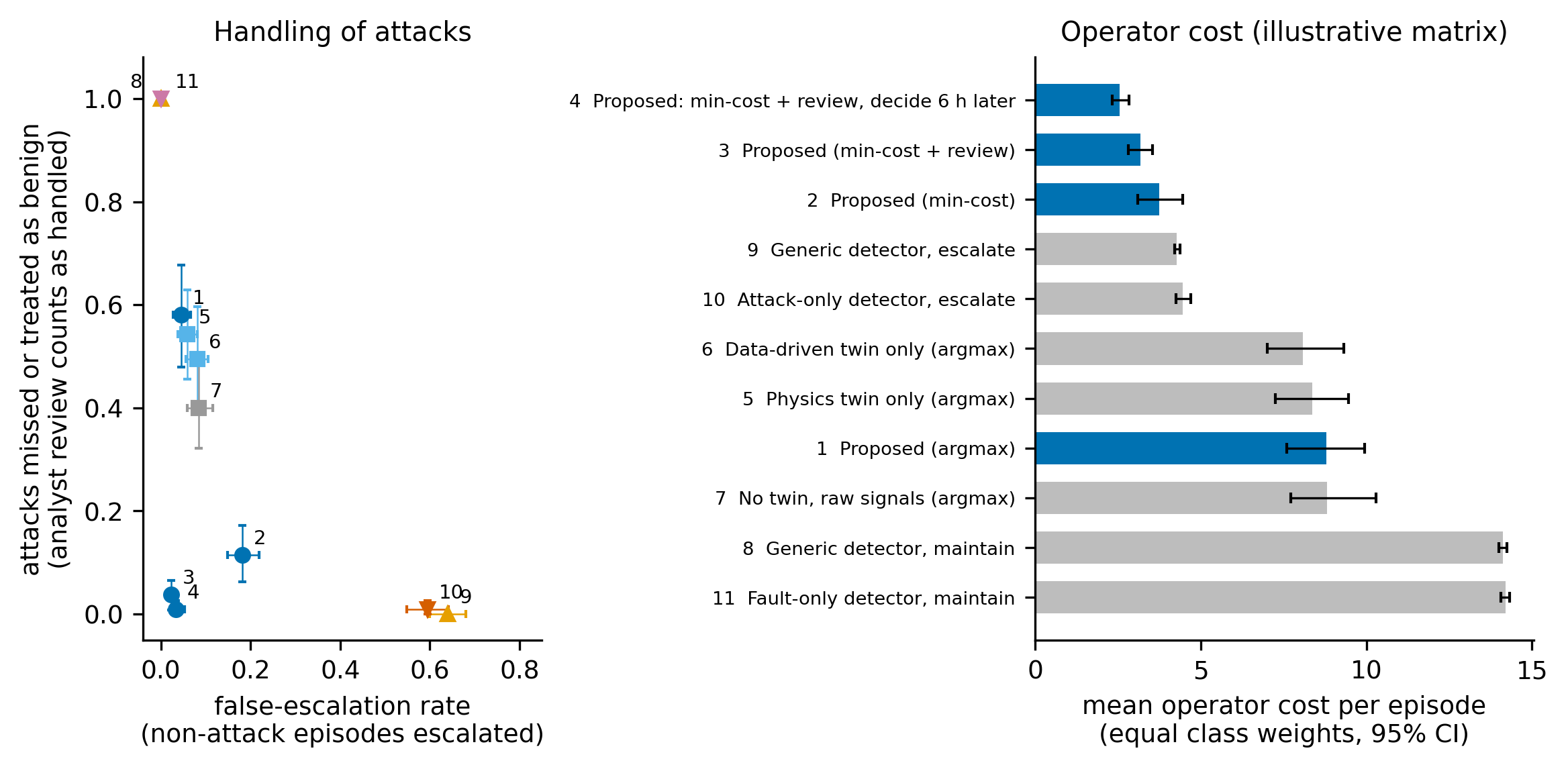}
\caption{Left: attack handling (numbers match the method list on the right). Right: mean operator cost with 95\% bootstrap CI under the illustrative cost matrix.}
\label{fig:cost}
\end{figure}

Three observations follow. First, argmax attribution is not useful as an action rule under these costs: its equal-weight cost (8.80 [7.59, 9.95]) is about twice that of escalating every alarm (4.28), because 58\% of attacks are labelled fault or environment at alarm time and treated as benign. Second, minimizing expected cost under the recalibrated posterior brings cost to 3.74 [3.10, 4.46], comparable to the escalate-everything baselines whose intervals it overlaps at equal weights; its advantage is clearer at the deployment prior (0.95 versus 1.59 and 1.54). Third, allowing analyst review lowers cost further to 3.18 [2.82, 3.54], and deciding six hours after the alarm to 2.56 [2.34, 2.83]. Under the review policy 55\% of anomalous episodes are sent to review, so this result depends on review being effective, which we test next.

\subsection{Sensitivity to the cost model}\label{sec:results-sens}

We recomputed expected cost over 144 settings: attack-miss cost 20--600, false-escalation cost 2--20, review effectiveness 0.5--1.0, and attack prior 0.5--10\%. A proposed variant had the lowest expected cost in every setting (the six-hour-deferral variant in 113, review at alarm time in 21, and plain minimum-cost in 10), including when review resolves only half of attacks. The margin is not always large. When false escalation is very cheap (cost 2 against a missed-attack cost of 60), escalating every alarm costs 0.65 versus 0.61 for the deferred policy, and attribution offers little. The six-hour deferral is not free: our cost model assigns no cost to waiting. Deciding at alarm time is preferable if an attack in progress inflicts more than about 0.23--0.50 cost units per hour of extra dwell time (roughly 0.4--0.8\% of the missed-attack cost per hour), which is a low bar.

\begin{figure}[ht]
\centering
\includegraphics[width=\linewidth]{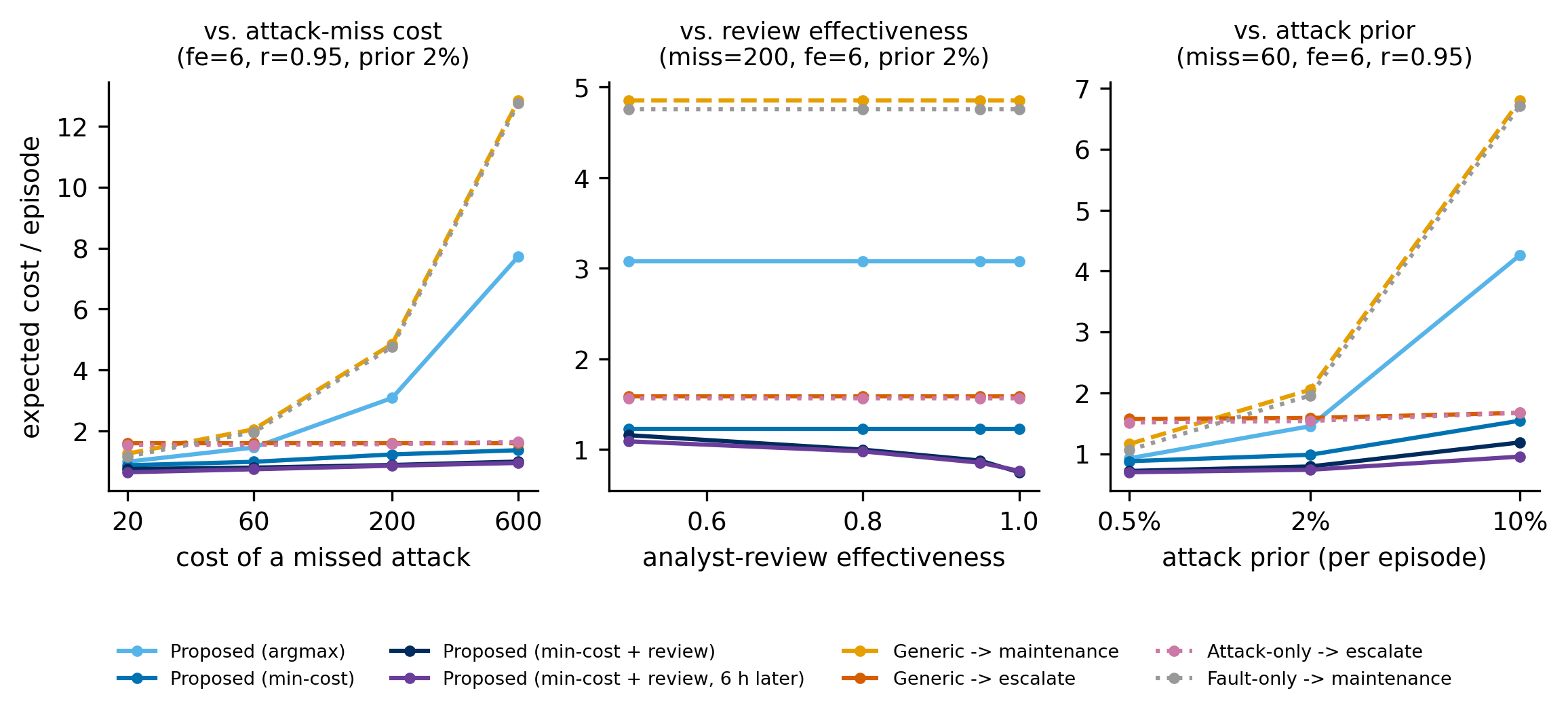}
\caption{Expected cost per episode (deployment-style prior) as attack-miss cost, analyst-review effectiveness and attack prior vary; other parameters as labelled above each panel.}
\label{fig:sensitivity}
\end{figure}

\subsection{Deviation magnitude}\label{sec:results-mag}

Scaling deviation magnitude from $0.25\times$ to $4\times$ the nominal distribution raises detection of faults from 65\% to 100\% and of environment effects from 67\% to 99\%. Attack detection stays at 96\%--100\% across the range. This should not be read as robustness to stealthy attacks: the multiplier does not scale replay attacks at all and mass-balance-preserving attacks remain large in absolute terms, so the sweep varies amplitude, not adversarial stealth. Attribution among detected episodes is not monotone in magnitude because small deviations are detected only when they are unusually distinctive (a selection effect); the joint rate of detecting and correctly attributing attacks rises from 17\% to 55\%.

\begin{figure}[ht]
\centering
\includegraphics[width=0.9\linewidth]{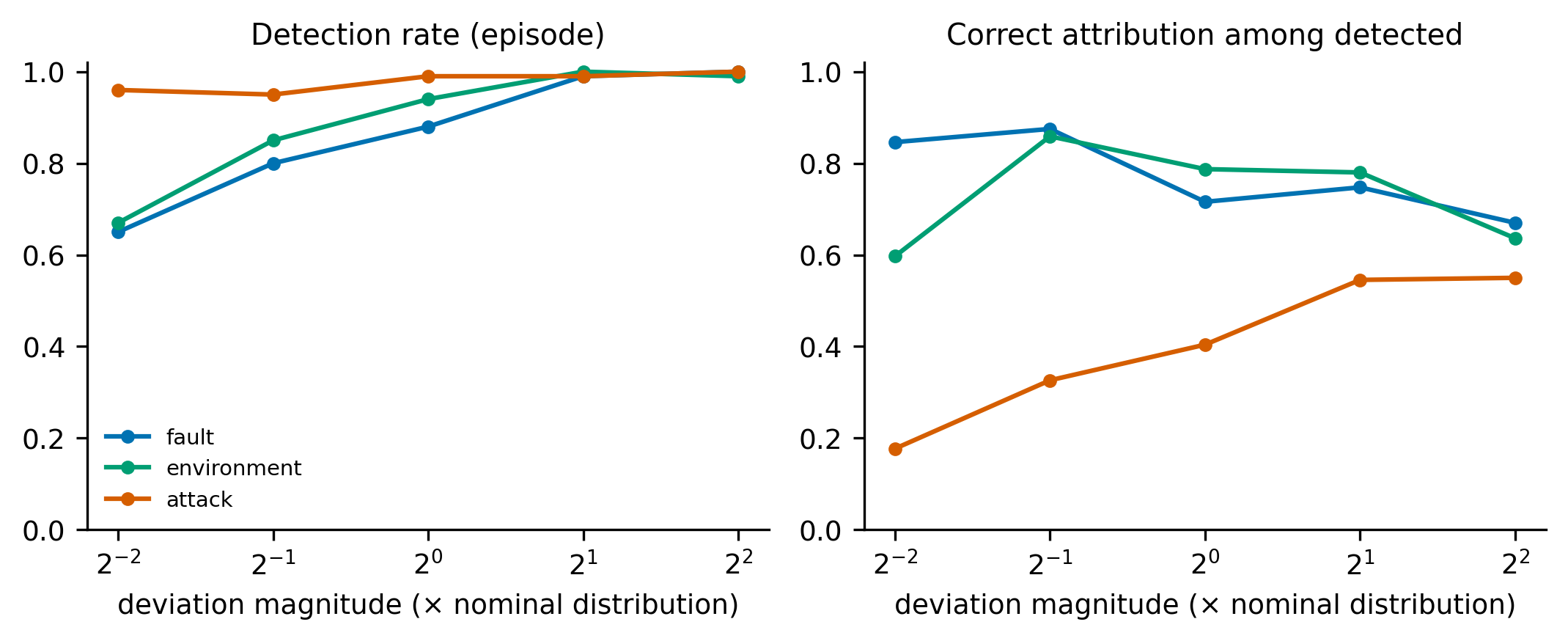}
\caption{Detection rate and correct attribution among detected episodes versus deviation magnitude, primary site, 100 episodes per cause and magnitude.}
\label{fig:magnitude}
\end{figure}

\subsection{New sites and stress tests}\label{sec:results-transfer}

Applying the primary-site pipeline unchanged to a new simulated site (different wells, a warmer climate and higher instrument noise) fails: macro-F1 0.15 and 0.55 false alarms per day. Re-identifying only the twin (physics parameters and residual statistics) on 60 normal episodes from the new site restores macro-F1 to 0.83 with 0.036 false alarms per day; recalibrating the alarm threshold as well changes little (0.027). A classifier retrained on the new site reached 0.79, but it used only 300 training episodes against 800, so this is not an upper bound. Transfer is easy here because every simulated site follows the same functional forms; real sites will violate them.

\begin{table}[ht]
\centering
\caption{Site transfer (300 test episodes at a new simulated site).}
\label{tab:transfer}
\small\setlength{\tabcolsep}{4pt}
\begin{tabular}{L{3.5cm}C{1.5cm}C{1.2cm}C{1.7cm}C{1.6cm}C{2.1cm}}
\toprule
\textbf{Configuration} & \textbf{Macro-F1} & \textbf{ECE} & \textbf{False alarms / day} & \textbf{Attack detected} & \textbf{Attack attributed at alarm} \\
\midrule
A. No adaptation        & 0.15 & 0.817 & 0.546 & 1.00 & 1.00 \\
B. Twin re-identified   & 0.83 & 0.069 & 0.036 & 0.99 & 0.45 \\
C. Twin + threshold re-set & 0.83 & 0.069 & 0.027 & 0.99 & 0.46 \\
D. Retrained on new site   & 0.79 & 0.044 & 0.024 & 0.99 & 0.43 \\
\bottomrule
\end{tabular}
\end{table}

In configuration A every alarm was labelled attack (attribution of faults and environment effects was 0.00 and 0.00), so its attack-attribution figure of 1.00 is an artifact of labelling everything as attack, not a success.

Stress tests use 100 attack episodes each (Table~\ref{tab:stress}; single run, no confidence intervals). Mass-balance-preserving and replay attacks are always detected, but attribution to attack at alarm time is only 68\% and 38\%. Removing the replay-fingerprint feature raised at-alarm attribution for both (to 80\% and 53\%), so the feature's window-level gain (Section~\ref{sec:results-abl}; replay recall on manifest windows is 0.93--0.96 across sites) does not carry over to alarm-time attribution of these attacks. A possible reason is that the fingerprint needs a window that is mostly replayed and so is weak at the first alarm; we did not test this. The twin-aware attacker was detected in 45\% of episodes against roughly 8\% expected by chance, but attributed to attack in 2\%. Detection here most plausibly reflects residual leakage because our attacker ignores thermal lag, not a designed defense.

\begin{table}[ht]
\centering
\caption{Attack stress tests, primary site.}
\label{tab:stress}
\small\setlength{\tabcolsep}{4pt}
\begin{tabular}{L{3.2cm}C{1.9cm}C{1.9cm}C{1.9cm}C{1.9cm}C{1.6cm}}
\toprule
\textbf{Attack} & \textbf{Detected (proposed)} & \textbf{Attributed to attack at alarm} & \textbf{Detected (no replay feature)} & \textbf{Attributed (no replay feature)} & \textbf{Chance detection} \\
\midrule
Mass-balance-preserving & 1.00 & 0.68 & 1.00 & 0.80 & 0.08 \\
Replay                  & 1.00 & 0.38 & 1.00 & 0.53 & 0.08 \\
Twin-aware (`full')     & 0.45 & 0.02 & 0.62 & 0.02 & 0.08 \\
\bottomrule
\end{tabular}
\end{table}

\begin{figure}[ht]
\centering
\includegraphics[width=\linewidth]{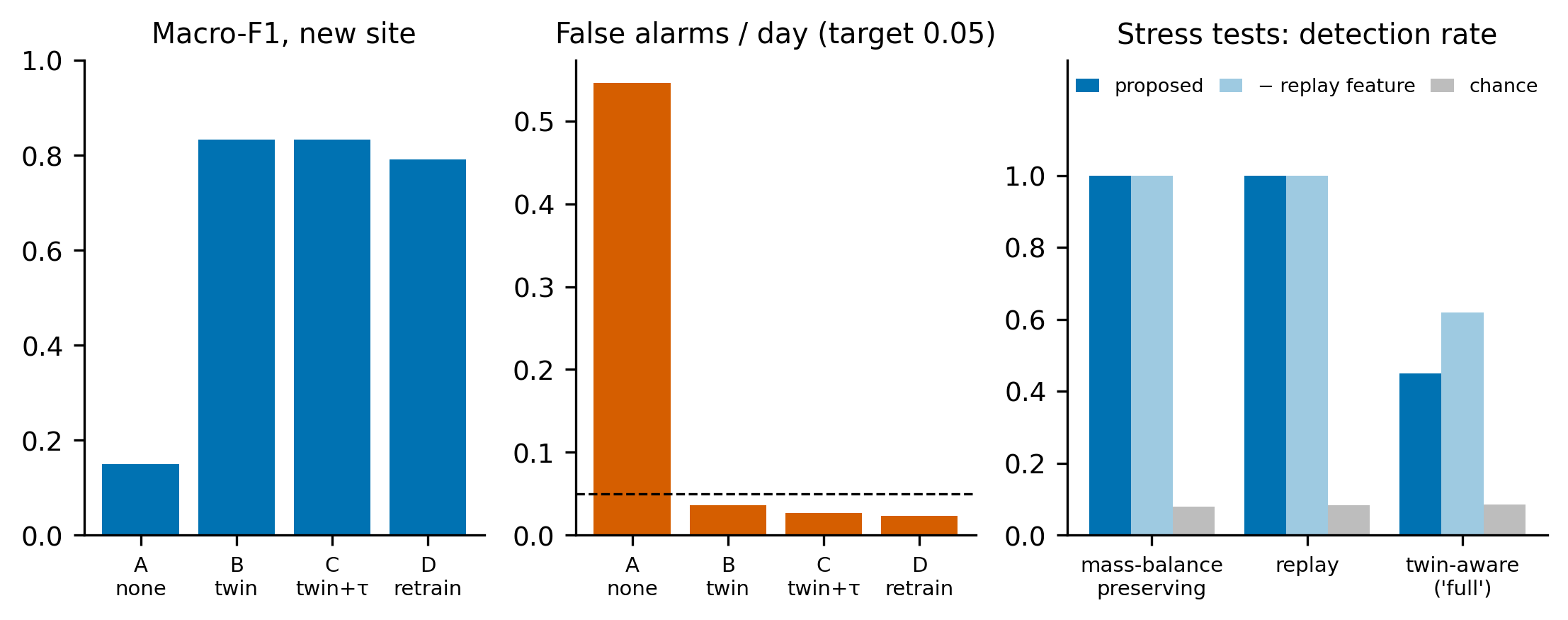}
\caption{Left and centre: transfer to a new simulated site under four adaptation levels (dashed line: false-alarm target). Right: detection rate under attack stress tests versus chance.}
\label{fig:transfer}
\end{figure}

\subsection{Illustrative episodes}\label{sec:results-episodes}

\begin{figure}[ht]
\centering
\includegraphics[width=0.92\linewidth]{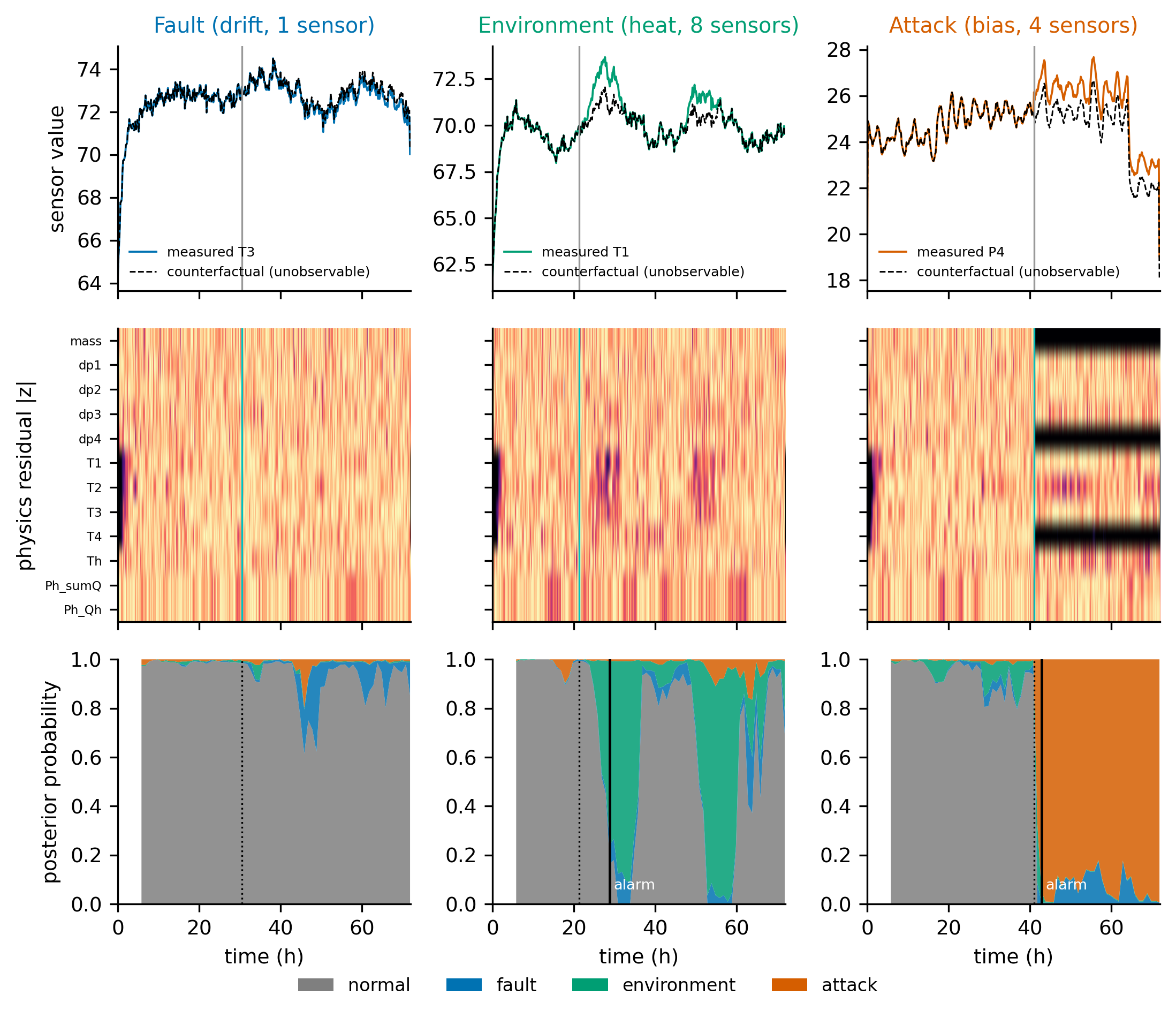}
\caption{One episode per cause (first random draws meeting a structural condition, not selected for performance). Top: measured signal and unobservable counterfactual. Middle: physics-residual magnitude by relation (start-up transient at left is a twin initialization effect). Bottom: attribution posterior; vertical black line marks the alarm. The fault example is a subtle drift that never triggers an alarm within 72~h.}
\label{fig:episodes}
\end{figure}

\section{Discussion}\label{sec:discussion}

What the evidence supports, within the simulator: a physics-informed twin identified from normal data supplies most of the information needed to separate causes; reasonably reliable posteriors (ECE 0.021--0.064) are usable for cost-aware decisions even though attribution at alarm time is often uncertain; and the practical value lies in deferral (analyst review or waiting) rather than in point predictions. Argmax attribution, taken literally, is worse than escalating everything under a cost structure where a missed attack is expensive.

What it does not support: that the data-driven twin or isotonic recalibration are worthwhile (they were not, here; topology features gave only a small gain of about 0.007 macro-F1); that attribution works on real instruments; that the cost advantage survives a real cost structure; or that a capable adversary can be attributed. The consistent results across three sites reflect three draws from one generative process, not three independent fields.

An operational reading is a two-stage response: raise the alarm quickly, treat early attribution as provisional, and commit to an attribution after the evidence has accumulated, with fault-versus-attack ambiguity routed to an analyst. Whether that trade-off is acceptable depends on the dwell-time cost of attacks, which the break-even analysis in Section~\ref{sec:results-sens} quantifies but which only operators can estimate.

\section{Limitations}\label{sec:limitations}

\begin{itemize}
\item Simulation only. Every result depends on the generative assumptions, and the simulator's overlap between single-sensor faults and attacks sets an identifiability floor that a real system may or may not share.
\item The cost matrix is illustrative and was chosen by the authors, not elicited from operators. Sensitivity analysis covers four of its parameters, not its structure.
\item The decision-time comparison ignores the cost of waiting; break-even values are reported instead.
\item Attackers are limited to false-data injection on sensor values via gateways or field devices. The twin-aware attacker ignores thermal dynamics and is not an optimal adversary. Adaptive attackers who observe the detector are not modelled.
\item The weather station is assumed trustworthy; spoofing it would defeat the weather-coupling features.
\item One process model, one topology, one sampling rate. The ARR structure is specific to a four-well pad.
\item Aggregate features were added after seeing prototype results on the same simulator; the manifest criterion excludes windows that no method could label, which applies to all methods alike but changes absolute numbers (results on all windows are reported).
\item Related-work characterizations rest on a brief literature search, and the claim that no prior work jointly attributes these three causes has not been established by a systematic review.
\end{itemize}

\section{Conclusion}\label{sec:conclusion}

On a simulated oilfield wellpad, a physics-informed twin plus probabilistic, cost-aware decision-making can separate sensor degradation, weather effects and attacks well enough to reduce operator cost relative to escalating every alarm, chiefly by deferring ambiguous cases. The approach cannot resolve single-sensor faults from single-sensor attacks that the simulator makes indistinguishable, and its behaviour under a knowledgeable attacker is untested beyond a deliberately imperfect stress test. The natural next step is validation on real historian data with labelled incidents; an adapter for that purpose accompanies the code.

\section*{Data and code availability}

Simulator, twins, feature extraction, experiments, analysis and figure code accompany this manuscript, with a script that reproduces all results. No field data were used.

\bibliographystyle{unsrt}
\bibliography{references}

\appendix

\section{Per-mode recall}\label{app:recall}

\begin{table}[ht]
\centering
\caption{Recall of the proposed classifier on manifest windows by injection mode, at each site.}
\label{tab:recall}
\small
\begin{tabular}{lcccc}
\toprule
\textbf{Mode} & \textbf{Site 1} & \textbf{Site 2} & \textbf{Site 3} & \textbf{Mean} \\
\midrule
Fault: stuck          & 1.00 & 1.00 & 1.00 & 1.00 \\
Fault: drift          & 0.48 & 0.46 & 0.51 & 0.48 \\
Fault: step           & 0.51 & 0.53 & 0.41 & 0.48 \\
Fault: gain           & 0.53 & 0.56 & 0.50 & 0.53 \\
Fault: noise bursts   & 0.48 & 0.64 & 0.50 & 0.54 \\
Environment: heat     & 0.68 & 0.75 & 0.76 & 0.73 \\
Environment: dust     & 0.78 & 0.78 & 0.79 & 0.78 \\
Environment: both     & 0.77 & 0.86 & 0.84 & 0.83 \\
Attack: bias          & 0.75 & 0.66 & 0.69 & 0.70 \\
Attack: ramp          & 0.51 & 0.64 & 0.55 & 0.57 \\
Attack: replay        & 0.96 & 0.96 & 0.93 & 0.95 \\
Attack: scale         & 0.74 & 0.84 & 0.68 & 0.76 \\
Attack: mass-balance  & 1.00 & 1.00 & 1.00 & 1.00 \\
\bottomrule
\end{tabular}
\end{table}

Recall is the fraction of manifest windows of that mode assigned to the correct cause (not the correct mode).

\end{document}